\documentclass[aps,przb,twocolumn,showpacs,letterpaper,superscriptaddress]{revtex4-1}
\usepackage{dcolumn}
\usepackage{bm}
\usepackage{xcolor}
\usepackage{multirow}
\usepackage{footnote}
\usepackage{tabularx}
\usepackage[utf8]{inputenc}
\usepackage[english]{babel}
\usepackage{amssymb}
\usepackage{amsmath}
\usepackage{graphicx}
\usepackage{color}
\usepackage{physics}
\usepackage{float}
\usepackage[normalem]{ulem}
\usepackage{booktabs}
\usepackage[colorlinks=true,linkcolor=blue,citecolor=blue,urlcolor=blue]{hyperref}
\usepackage[capitalize,nameinlink,noabbrev]{cleveref}

\renewcommand{\vec}[1]{\boldsymbol{#1}}
\graphicspath{{Figs/}}
\setcitestyle{square,numbers} 
 
\newcommand{\be}{\begin{equation}}
\newcommand{\ee}{\end{equation}}
\newcommand{\bea}{\begin{eqnarray}}
\newcommand{\eea}{\end{eqnarray}}

\newcommand{\bt}{\begin{table}}
\newcommand{\et}{\end{table}}

\begin{document}

\title{Twistronic control of shift current in multilayer moiré system}

\author{Michele Bagaglini}
\affiliation{Department of Physical and Chemical Sciences, University of L{'}Aquila, Via Vetoio, 67100 L{'}Aquila, Italy}
\affiliation{CNR-SPIN L{'}Aquila, Via Vetoio, 67100 L{’}Aquila, Italy}

\author{Cesare Tresca}
\affiliation{CNR-SPIN L{'}Aquila, Via Vetoio, 67100 L{’}Aquila, Italy}

\author{Federico Bisti}
\affiliation{Department of Physical and Chemical Sciences, University of L{'}Aquila, Via Vetoio, 67100 L{'}Aquila, Italy}

\author{Gianni Profeta}
\affiliation{Department of Physical and Chemical Sciences, University of L{'}Aquila, Via Vetoio, 67100 L{'}Aquila, Italy}
\affiliation{CNR-SPIN L{'}Aquila, Via Vetoio, 67100 L{’}Aquila, Italy}

\begin{abstract}
The bulk photovoltaic effect in non-centrosymmetric materials provides an alternative mechanism for the conversion of light into a current response compared to p–n junctions. Among its various contributions, the shift current is particularly attractive because it is governed by the geometric properties of electronic wavefunctions and can generate large photocurrents in low-dimensional materials. Here, we investigate the evolution of the shift current response in mono-, bi-, and trilayer H-MoS$_2$, as well as in twisted moiré bilayers and trilayers. To describe large moiré supercells we develop a Slater–Koster tight-binding model parametrized from first-principles calculations. The resulting electronic structures and shift-current responses are compared with density functional theory calculations and Wannier-interpolated results to verify the accuracy of the approach. The model accurately reproduces the electronic structure near the band edges and captures the main spectral features of the shift current conductivity. We show that twisting breaks the crystal symmetry and activates additional conductivity tensor components that are forbidden in untwisted structures, leading to new tunable in-plane photocurrent components. Analysis of the shift distance reveals a direct connection between the twist-induced modification of the electronic wavefunctions and the increase of the nonlinear response. Our results establish the twist angle as an effective parameter for engineering shift current generation in multilayer transition-metal dichalcogenide base systems and demonstrate that tight-binding approaches provide a practical route for exploring nonlinear optical phenomena in large-scale moiré materials beyond the limits of conventional first-principles calculations.

\end{abstract}

\maketitle

\section{Introduction}

One of the main objectives of research in the field of sustainable energy is to increase the production of energy from renewable energy sources. Photovoltaic solar energy plays a central role due to its abundance, scalability, and rapidly decreasing production cost \citep{Green2025}. 
In recent years, research has increasingly focused on improving the efficiency and functionality of photovoltaic devices by exploring novel materials, unconventional physical mechanisms for photocurrent generation, and strategies to overcome fundamental efficiency limits. In particular, significant efforts are directed toward low-dimensional materials \citep{sauer2023shift}\citep{esteve2025excitons}\citep{ye2023manipulation}\citep{habara2023symmetry}, ferroelectrics \citep{Young2012}\citep{dai2021first}, and topological systems \citep{Ogawa2016}\citep{Braun2016}, where new mechanisms beyond the conventional photovoltaic effect can emerge.

In traditional solar cells the $p$-$n$ junction creates an internal electric field that drives the electron-hole separation process, enabling the photovoltaic effect to convert light into current. Such junctions suffer from intrinsic limitations, such as the maximum achievable conversion efficiency given by the Shockley–Queisser limit \citep{shockley2018detailed}, and energy losses due to carrier thermalization and drift processes \citep{tan2016shift}.

An alternative to $p$-$n$ junctions is represented by materials that exhibit the bulk photovoltaic effect (BPVE) \citep{tan2016shift}\citep{cook2017design}, a nonlinear optical response that arises in systems with an inversion symmetry breaking. Unlike conventional photovoltaic mechanisms, the BPVE does not rely on built-in electric fields, but instead originates from intrinsic properties of the materials.

Among the different contributions to the BPVE, the shift current has been identified as a dominant mechanism, with a direct connection to the geometric and topological properties of the electronic wavefunctions, such as the Berry connection \citep{sipe2000second}\citep{Morimoto2016}.

The shift current (SC) originates from the displacement of the charge center when an electron is promoted from the valence band to the conduction band \citep{sipe2000second}. Unlike $p$-$n$ junctions, it is possible to overcome the Shockley–Queisser limit because the photovoltage can exceed the band gap \citep{tan2016shift}\citep{spanier2016power}. In addition, there is a minimization of the energy loss due to drifting processes, because the photocurrent is generated by the coherent evolution of the electron and hole wavefunctions, and the entire volume of the material participates in the generation of the current.

In recent years, research has investigated two-dimensional (2D) materials, which provide an ideal platform for next-generation optoelectronic and photovoltaic devices \citep{Wang2012}. Their reduced dimension is suitable for the integration into electronic circuits \citep{Wang2012}\citep{Lee2014}, and while also allowing for strong tunability through external perturbations such as strain \citep{Yu2024} and electric fields \citep{Novoselov2005}. Moreover, the weak van der Waals interaction between layers enables the fabrication of heterostructures offering a large flexibility in material design \citep{Liang2019}, for example the possibility of engineering their electronic and optical properties via stacking and twisting. When two layers are combined with a relative twist angle, a moiré superlattice emerges, leading also to the modification of electronic structures. This so-called ''twistronics'' approach has enabled the observation of a wide range of phenomena, including metal-insulator transition, superconductivity, and non-trivial topological phases \citep{Cao2018}\citep{Devakul2021}\citep{MoralesDurn2024}. 

These features make twisted 2D systems a powerful platform for band structure engineering and for controlling topological and geometric properties of electronic states. Such control impacts nonlinear optical responses, including the bulk photovoltaic effect, opening new possibility for the shift current generation.

Among the various families of 2D materials, transition metal dichalcogenides (TMDs) have emerged as promising candidates for photovoltaic and nonlinear optical applications \citep{Li2018}. Monolayer TMDs exhibit a band gap in the visible range, strong light-matter interaction, and large spin–orbit coupling, making them highly efficient for optoelectronic processes.

Furthermore, TMD multilayers and twisted heterostructures provide additional degrees of freedom to engineer symmetry breaking and electronic structure. These characteristics make these systems particularly suitable candidates for investigating and optimizing the shift current contribution to the BPVE.

In this work, we investigate the shift current response in $H$-MoS$_2$ based systems whose shift current conductivity exhibits pronounced peaks in the solar energy spectrum range relevant for photovoltaic applications. Previous theoretical studies have analyzed the monolayer \citep{esteve2025excitons}\citep{dai2021first}\citep{cheng2025shift} and the untwisted bilayer \citep{xiao2022non}, while experimental evidence has also been reported for twisted bilayers \citep{huang2025phonon}.
Moreover, to ensure an effective response to unpolarized sunlight, a highly anisotropic material must be employed \citep{cook2017design}. For this reason, in addition to the untwisted multilayer configurations, we analyze twisted systems. Studying how these changes affect the shift current conductivity tensor as a function of the twisting angle.

In principle, an $ab$ $initio$ approach provides higher accuracy in the description of electronic properties and optical response within density functional theory (DFT). In particular, first-principles calculations combined with Wannier interpolation enable a direct evaluation of the nonlinear optical conductivity tensors associated with the shift current \citep{IbaezAzpiroz2018}. Another general method for evaluating the shift current conductivity is based on localized Gaussian basis sets, which allow to describe the optical matrix elements within an $ab$ $initio$ framework \citep{GarcaBlzquez2023}. However, for twisted systems, DFT becomes computationally prohibitive due to the large size of the moiré supercells, which can contain thousands of atoms even for relatively small twist angles.

For this reason, we adopt a tight-binding (TB) approach with a Slater-Koster (SK) parametrization \citep{slater1954simplified}, which provides an efficient and flexible framework to describe large-scale systems while retaining the essential features of the electronic structure. In particular, tight-binding models allow for an accurate interpolation of the band structure and optical matrix elements, enabling the evaluation of nonlinear response functions with significantly reduced computational effort \citep{Koshino2018}.

We show that, when properly parametrized, the tight-binding model captures the key features of the nonlinear shift current response, including the spectral position and intensity of the main peaks, as well as the anisotropic behavior of the conductivity tensor. This approach therefore allows for a reliable and systematic analysis of the shift current in both untwisted and twisted multilayer systems, where large-scale simulations are otherwise inaccessible within standard first-principles methods.

\section{Methods}

$DFT\quad calculations.-$ The density functional theory (DFT) calculations for monolayer, bilayer, and trilayer $MoS_2$ are performed using the QUANTUM ESPRESSO package \citep{giannozzi2009quantum}\citep{giannozzi2017advanced}. A vacuum spacing of about $\sim$20 \AA~is introduced between periodic replicas to minimize the interlayer interactions. To better approximate an isolated system, we also apply a truncation of the Coulomb interaction along the $\hat{z}$ direction \citep{sohier2017density}.
Exchange-correlation effects are treated within the Perdew–Burke–Ernzerhof (PBE) functional \citep{hamann2013optimized}. The self-consistent calculations are carried out on a $12\times12\times1$ k-point grid, using a plane-wave cutoff of 100 Ry and a Gaussian smearing of 0.001 Ry. The lattice constant of the monolayer is set to $a = 3.18$ \AA, in agreement with the literature \citep{venkateswarlu2020electronic}. The wavefunctions are then computed on the same k-point grid.
While $ab$ $initio$ calculations typically underestimate the electronic gap, our result of $\Delta_{DFT} \simeq 1.66$ eV is sufficiently close to the experimental optical gap of $\Delta_{exp} \simeq 1.8 - 1.9$ eV \citep{Mak2010} for the purposes of our study. Therefore, no further corrections, such as a scissor operator, are required to achieve a reliable description of the system's response.
A set of 11 maximally localized Wannier functions, for every single layer, is constructed using Mo $4d$ and S $3p$ orbitals as initial projectors, and obtained with the WANNIER90 package \citep{mostofi2014updated}. For the bilayer and trilayer systems, we use an interlayer distance of $6.80$ \AA, consistent with previous studies \citep{venkateswarlu2020electronic}.\\

$TB\quad model.-$ We develop a Slater-Koster (SK) parameterization for the tight-binding model through the python package TBPLaS \citep{li2023tbplas}. The SK parameters are fitted on the top of DFT with particular focus on the top valence band and first conduction band.
For the intralayer terms, we include hopping between first nearest neighbors, namely Mo–Mo, Mo–S, and S–S atoms in different unit cells, as well as S–S intracell interactions (see Fig.\ref{fig:monolayer_intraparameters}).
For the interlayer SK parameters in the bilayer and trilayer system we include only first neighbours interactions: Mo-Mo, Mo-S and S-S between the different layers.
We reproduce the modification with the distance of the interlayer SK terms with an exponential scaling rule in line with material in the family of TMD \citep{fang2015ab} and twisted bilayer graphene \citep{trambly2012numerical}\citep{trambly2010localization}:

\be
\label{eq:scaling_rule}
V_i(d) = V^0_i exp\Bigg(-q_i \frac{d-d^0_i}{d^0_i}\Bigg)F_c(d)\quad.
\ee
Where $V_i^0$ are the SK parameters given in Tab.\ref{tab:interlayer_MoS2}, $d^0_i$ is the interatomic distance before twisting between the $i$-th couple of neighbors, $d$ is the new distance after the twisting, $q_i$ is a fixed coefficient that reduces a factor of 10 the value of the parameters between first and second neighbor hopping terms \citep{trambly2012numerical}:

\be
q_i = \frac{\sqrt{3}ln(10)d^0_i}{(\sqrt{3}-1)a} \quad.
\ee

The $F_c(d)$ coefficient is a smooth cutoff function \citep{mehl1996applications}:

\be
F_c(d) = \Big(1 + exp\Big(\frac{d - r_c}{l_c}\Big)\Big)^{-1}
\ee

where $r_c$ is the cutoff distance, $r_c = 2.5a = 7.95$ \AA~\citep{mehl1996applications}, and $l_c = 0.265$ \AA. For $d \ll r_c$, $F_c(d) \simeq 1$ and for $d \gg r_c$, $F_c(d) \simeq 0$.\\

To build commensurate moiré cells we use the rules for twisted heterostructures bilayer system with triangular unit cell \citep{trambly2010localization}. The possible twisting angles for commensuration are defined by two integers $n$ and $m$ such that: 

\be
\label{eq:moire_angle}
cos\theta = \frac{n^2+4nm+m^2}{2(n^2+nm+m^2)}
\ee

where $n$ and $m$ are the integers that define the moiré lattice vectors $\vec{t}_1 = n\vec{a}_1 + m\vec{a}_2$ and $\vec{t}_2 = -m\vec{a}_1 + (n+m)\vec{a}_2$, with $\vec{a}_1$ and $\vec{a}_2$ the monolayer lattice vectors. We consider the following pair of ($n$; $m$): (1; 2), (2; 3), (3; 4) and (4; 5), corresponding to the twisting angles (deg): 21.787, 13.174, 9.430 and 7.341.\\

$Shift\quad Current\quad calculation.-$ The Shift Current (SC) is calculated via the postW90 package \citep{mostofi2014updated} for the data obtained via DFT and via the Python package WannierBerri \citep{tsirkin2021high} for the TB Hamiltonian. We use a $150\times150\times1$ k point grid and a smearing of 0.005 eV for the calculation of the SC. The definition of the SC is \citep{sipe2000second}\citep{ibanez2018ab}:

\be
\label{eq:J_sc}
J_{sc}^{abc}(0; \omega; -\omega) = \sigma_{sc}^{abc}(0; \omega; -\omega)E_{b}E_{c}
\ee

where the $a$-index is the Cartesian direction of the density-current, $b$ and $c$ are Cartesian directions of the linearly polarized electric fields $E$, and the tensor $\sigma_{sc}^{abc}$ is defined as:

\begin{widetext}
\be
\label{eq:sigma_sc}
\sigma_{sc}^{abc}(0; \omega, -\omega) = -\frac{i\pi e^3}{4\hbar^2}\sum_{\vec{k}}\sum_{n,m}(f_{m\vec{k}}-f_{n\vec{k}})(r_{mn}^br_{nm}^{c;a} + r_{mn}^cr_{nm}^{b;a})(\delta(\omega_{mn}-\omega)+\delta(\omega_{nm}+\omega))
\ee  
\end{widetext}

where $f_{m\vec{k}}$ the Fermi-Dirac occupation of the band $m$, the $b^{th}$ component of the dipole matrix element $r_{nm}^b = i\bra{u_n}\partial_{k_b}\ket{u_m} = A_{nm}^b$ is also the interband Berry connection. The term $r_{nm}^{c;a} = \partial_a r^c_{nm} - i(A^a_{nn}-A^a_{mm})r^{c}_{nm}$ is the generalized derivative of the dipole matrix element.

Since the systems analyzed are two-dimensional, the 3D SC conductivities are rescaled to 2D as $\sigma^{2D}_{sc} = \sigma^{3D}_{sc} \cdot c$, as mentioned in literature \citep{strasser2022nonlinear}, where $c$ represents the out-of-plane unit cell dimension; specifically, $c = 20$ \AA~ for the monolayer, and $c = 40$ \AA~ for the bi- and trilayers.\\

We neglect spin–orbit coupling (SOC) in all our calculations since previous works \citep{cheng2025shift} have shown that its effect on the MoS$_2$ shift current profile is negligible.

\section{Parametrization of $\text{MoS}_2$}

In this section we study the non-twisted structure of three different systems (mono-, bi-, and trilayer) in order to obtain the parameters for the TB model and to calculate the shift current (SC) as a function of the number of layers. For this purpose, the section is divided into three subsections, each corresponding to a different number of layers.

In the monolayer case, we study the electronic band structure using DFT calculations and we fit the onsite and intralayer hopping parameters of the Slater-Koster tight-binding (SKTB) model from the the Wannier interpolation on DFT data. In the bilayer subsection, we determine the interlayer parameters. In the trilayer discussion, no additional fitting is required, since the previously obtained parameters are sufficient to reproduce the DFT band structure.

In each subsection, we compare the shift current calculated from $ab$ $initio$ methods using postW90 with that obtained from the SKTB model using Wannierberri package.

\subsection{Monolayer}

\begin{figure}[h]
    \centering
    \includegraphics[width=1\linewidth]{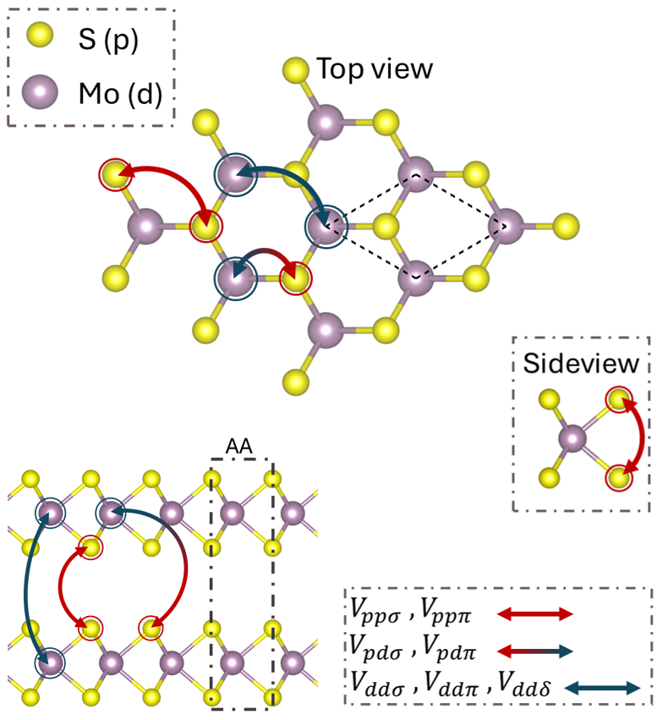}
    \caption{The $H$-MoS$_2$ structure. The three atoms in the unit cell have fractional coordinates given by: Mo = (0,0,0), S = (1/3,1/3,$\pm z_S$/c), where $z_S$ = 1.56 \AA. The lattice constant is $a$ = 3.18 \AA. The crystal is spanned by the two lattice vectors: $t_1 = a(\sqrt{3}/2,-1/2,0)$ and $t_2 = a(\sqrt{3}/2,1/2,0)$. In the bilayer configuration, with an AA stacking, the distance between the layers is 6.80 \AA. In this configuration all the atoms in the unit cell lie in the xz plane. The different coloured arrows indicate the SK parameters for the hoppings between the different combinations of Mo (blue arrows) and S (red arrows).}
    \label{fig:monolayer_intraparameters}
\end{figure}

The monolayer $H$-MoS$_2$ has a non-centrosymmetric structure \citep{cheng2025shift} that is shown in Fig.\ref{fig:monolayer_intraparameters}.  

\begin{figure}
    \centering
    \includegraphics[width=1\linewidth]{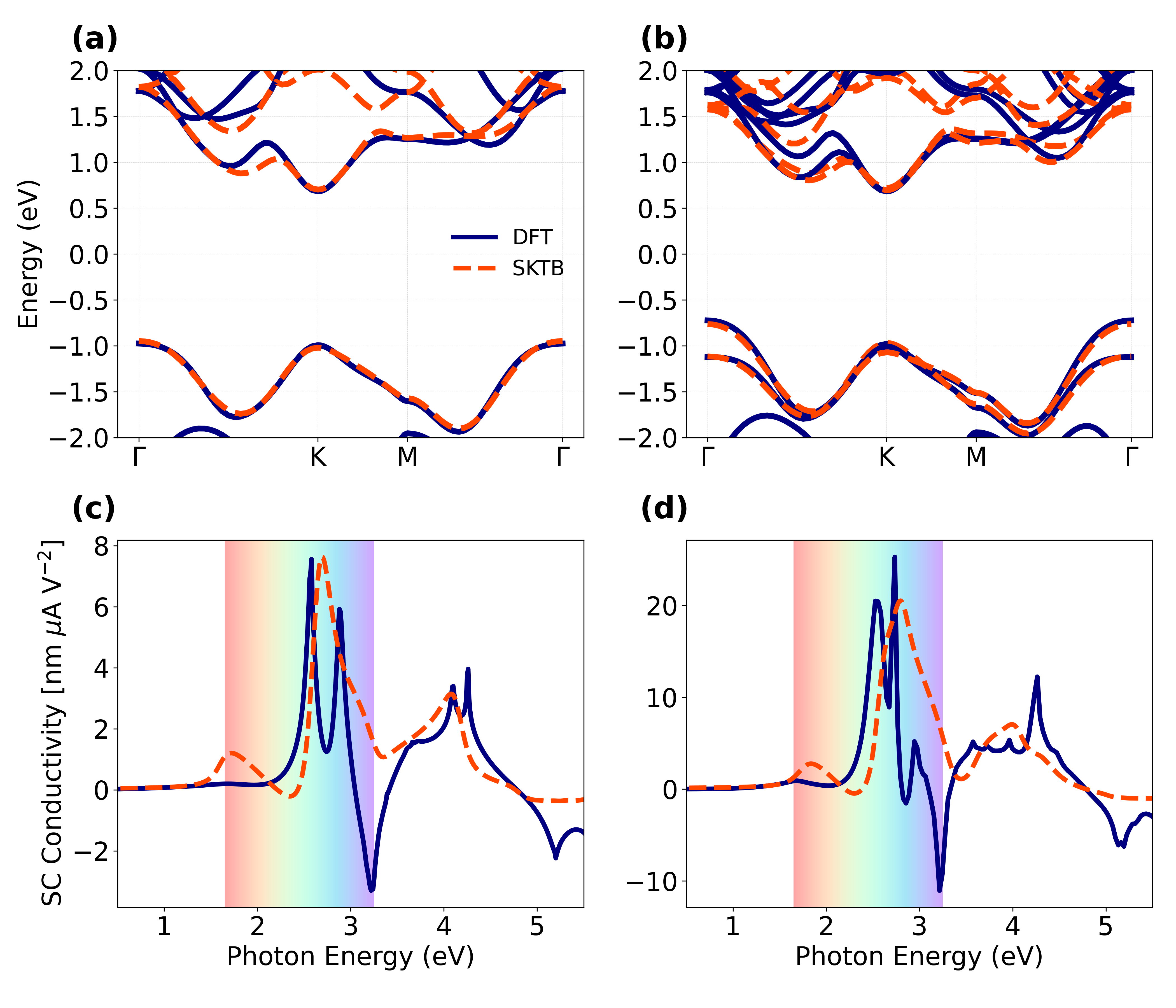}
    \caption{DFT band structure (blue solid line) of monolayer $H$-MoS$_2$ (panel $a$) and bilayer $H$-MoS$_2$ (panel $b$) and band structure obtained from the fitted SKTB model (dashed orange line). In panel $c$ and $d$ the comparison between the shift current calculated within the $ab$ $initio$ framework (blue solid line) and that obtained from the tight-binding model (dashed orange line).}
    \label{fig:mono-bi}
\end{figure}

From the Wannier bands we fit the onsite energies and the intralayer hopping, refer to the caption of Fig.\ref{fig:monolayer_intraparameters} for details of the hopping parameters. The bands obtained by fit are shown in Fig.\ref{fig:mono-bi}, panel $a$, compared to Wannierized bands, which are referred to as the DFT data. The parameters obtained for the onsite energies and hopping parameters are presented in Tab.\ref{tab:intralayer_MoS2}.\\

The resulting SKTB bands exhibit an excellent agreement with the DFT results, particularly in describing the valence band maximum and the conduction band minimum, showing a perfect match at the high-symmetry points of the Brillouin zone.

\begin{table}[h]
\centering
\small
\renewcommand{\arraystretch}{1.3}
\begin{tabular}{|c|l|c|l|}
    \hline
    \textbf{At.} & \textbf{Onsite (eV)} & \textbf{First N.} & \textbf{Hop. (eV)} \\
    \hline
    \hline
    & $4d_{z^2} = -1.189$ & & $V_{dd\sigma}=-0.701$ \\
    & $4d_{xy} = -0.817$ & Mo & $V_{dd\pi}=0.534$ \\
    Mo & $4d_{x^2-y^2} = -0.817$ & & $V_{dd\delta}=0.198$ \\
    \cline{3-4}
    & $4d_{yz} = -0.977$ & & $V_{dp\sigma}=-5.469$ \\
    & $4d_{zx} = -0.977$ & S & $V_{dp\pi}=2.274$ \\
    & & & \\
    \hline
    & $3p_x = -22.36$ & S & $V_{pp\sigma}=3.064$ \\
    S & $3p_y = -22.36$ & (Intra) & $V_{pp\pi}=7.242$ \\
    \cline{3-4}
    & $3p_z = -25.08$ & S & $V_{pp\sigma}=3.434$ \\
    & & (Inter) & $V_{pp\pi}=-3.275$ \\
    \hline
\end{tabular}
\caption{Fitted parameters for the monolayer $1H$-MoS$_2$: onsite parameters for the orbital energies and Slater-Koster tight-binding intralayer hopping parameters between first neighbors, for the sulfide we distinguish between S-S intra unit cell and inter unit cell interactions.}
\label{tab:intralayer_MoS2}
\end{table}

Due to the symmetry of the system the only non-zero SC-components are $-\sigma^{xxx}_{sc} = \sigma^{xyx}_{sc} = \sigma^{yxy}_{sc} = \sigma^{yyx}_{sc}$. In Fig.\ref{fig:mono-bi} panel $c$, we show the comparison between SC obtained via DFT and TB. The shift current calculated from DFT is comparable to the values reported in previous works \citep{cheng2025shift}\citep{ghosh2025choosing}. We are interested in the shift current response in the solar spectrum range, where the two main positive peaks from the $ab$ $initio$ calculation are observed. Using the SKTB model, we are able to reproduce these features with a single broader peak. The difference in the wavefunctions obtained from the two approaches explains why a complete quantitative agreement cannot be achieved.
Within this framework, the SKTB results are in good qualitative agreement with the DFT calculations, both in the solar range and outside it.

\subsection{Bilayer}

We start considering a AA configuration for the bilayer $2H$-MoS$_2$, with the same monolayer symmetry P-6m2, where the Mo (S) atoms of top layer lie above the other Mo (S) atoms of the bottom layer. In this configuration the crystal is still non-centrosymmetric, Fig.\ref{fig:monolayer_intraparameters}.

We obtain the SK-parameters for the interlayer hopping terms, Tab.\ref{tab:interlayer_MoS2}, fitting the bands obtained via an interpolation of the Wannier90 bands. The Wannier interpolated bands, referred as DFT, and SKTB bands are shown in Fig.\ref{fig:mono-bi} panel $b$.\\

\begin{table}[h]
\centering
\small
\renewcommand{\arraystretch}{1.3}
\begin{tabular}{|c|c|l|}
    \hline
    \textbf{Atom} & \textbf{Neighbor} & \textbf{Hop. (eV)} \\
    \hline
    \hline
    Mo & Mo & \begin{tabular}[c]{@{}l@{}} $V_{dd\sigma}=-0.0358$ \\ $V_{dd\pi}=-0.310$ \\ $V_{dd\delta}=0.0475$ \end{tabular} \\
    \hline
    Mo & S & \begin{tabular}[c]{@{}l@{}} $V_{dp\sigma}=-0.403$ \\ $V_{dp\pi}=-0.0636$ \end{tabular} \\
    \hline
    S & S & \begin{tabular}[c]{@{}l@{}} $V_{pp\sigma}=1.223$ \\ $V_{pp\pi}=-2.411$ \end{tabular} \\
    \hline
\end{tabular}
\caption{Fitted tight-binding interlayer hopping parameters for bilayer $2H$-MoS$_2$.}
\label{tab:interlayer_MoS2}
\end{table}

In this configuration all the atoms in the unit cell lie on the xz plane, so the $\sigma^{yyy}_{sc}$ component of the shift current is null as in the monolayer configuration. The shift current is shown in Fig.\ref{fig:mono-bi} panel $d$. The agreement between the two approaches in the energy range in consideration is good. Similar to the monolayer case, we reproduce the two main peaks with a single broader peak of comparable magnitude.

\subsection{Trilayer \textbf{$MoS_2$}}

Following the analysis of monolayer and bilayer MoS$_2$, we extend our study to trilayer systems. We consider an AAA-stacked trilayer configuration. In this stacking, the crystal preserves the same point group symmetry (P-6m2) as the monolayer and bilayer cases.

\begin{figure}[h]
    \centering
    \includegraphics[width=0.95\linewidth]{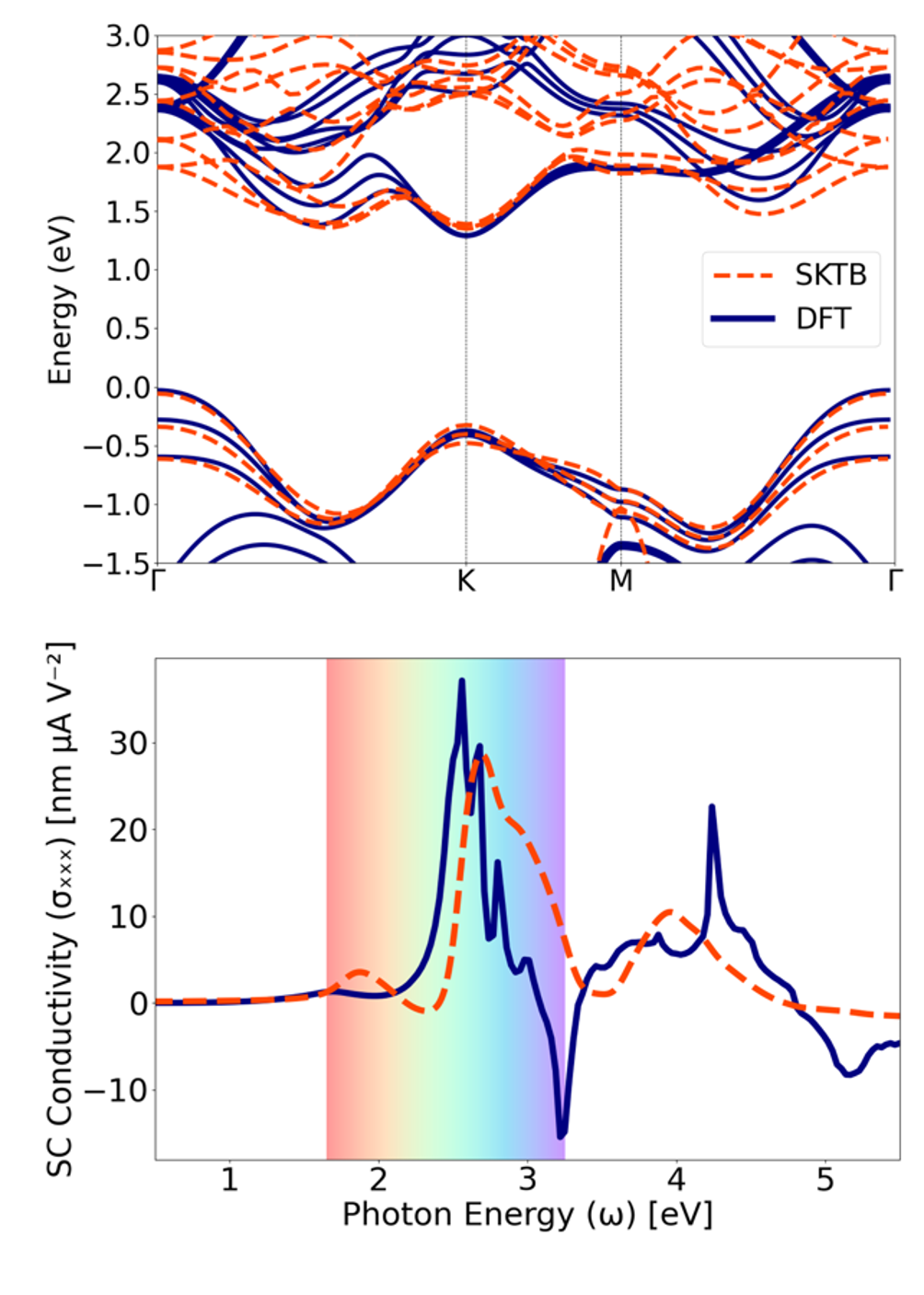}
    \caption{DFT band structure of trilayer MoS$_2$ (blue, top panel) and band structure obtained from the fitted SKTB model (orange, top panel). Comparison between the shift current calculated within the $ab$ $initio$ framework (blue, bottom panel) and that obtained from the tight-binding model (orange, bottom panel).}
    \label{fig:bands+sc-tri}
\end{figure}

Using the onsite, intralayer, and interlayer SKTB parameters fitted for the monolayer and bilayer systems (Tab.\ref{tab:intralayer_MoS2} and \ref{tab:interlayer_MoS2}), we construct the trilayer Hamiltonian without performing any additional fitting on first-principles data. Despite this, the resulting band structure near the Fermi level is in good agreement with the DFT results, as shown in Fig.\ref{fig:bands+sc-tri}. This demonstrates the transferability of the tight-binding parametrization to multilayer systems.

As in the previous configurations the tight-binding model is able to successfully reproduce the main features of the shift current, confirming its suitability for the study of multilayer systems, Fig.\ref{fig:bands+sc-tri}.\\

Moreover, we observe a trend in the shift current calculations, both in the DFT and SKTB results, regarding the magnitude of the main peak: the maximum value of the shift current is approximately proportional to the number of layers with respect to the monolayer case. This behavior suggests that the interlayer coupling affects the positions of the charge centers, leading to an overall enhancement of the shift current response.

\section{Twisting}

We now consider the twisted bilayer systems, using Eq.\ref{eq:moire_angle} to obtain different twisting angles for commensurate moiré cells. To account for the variation of the interlayer hopping terms due to the twisted geometry, we consider the scaling of the parameters with the distance relative to the AA configuration, by using Eq.\ref{eq:scaling_rule}.\\
We test the quality of this model through the comparison with the DFT bands structure  calculated for the smallest moiré cell corresponding to a twisting angle of $\theta = 27.787^\circ$.

\begin{figure}[h]
    \centering
    \includegraphics[width=1\linewidth]{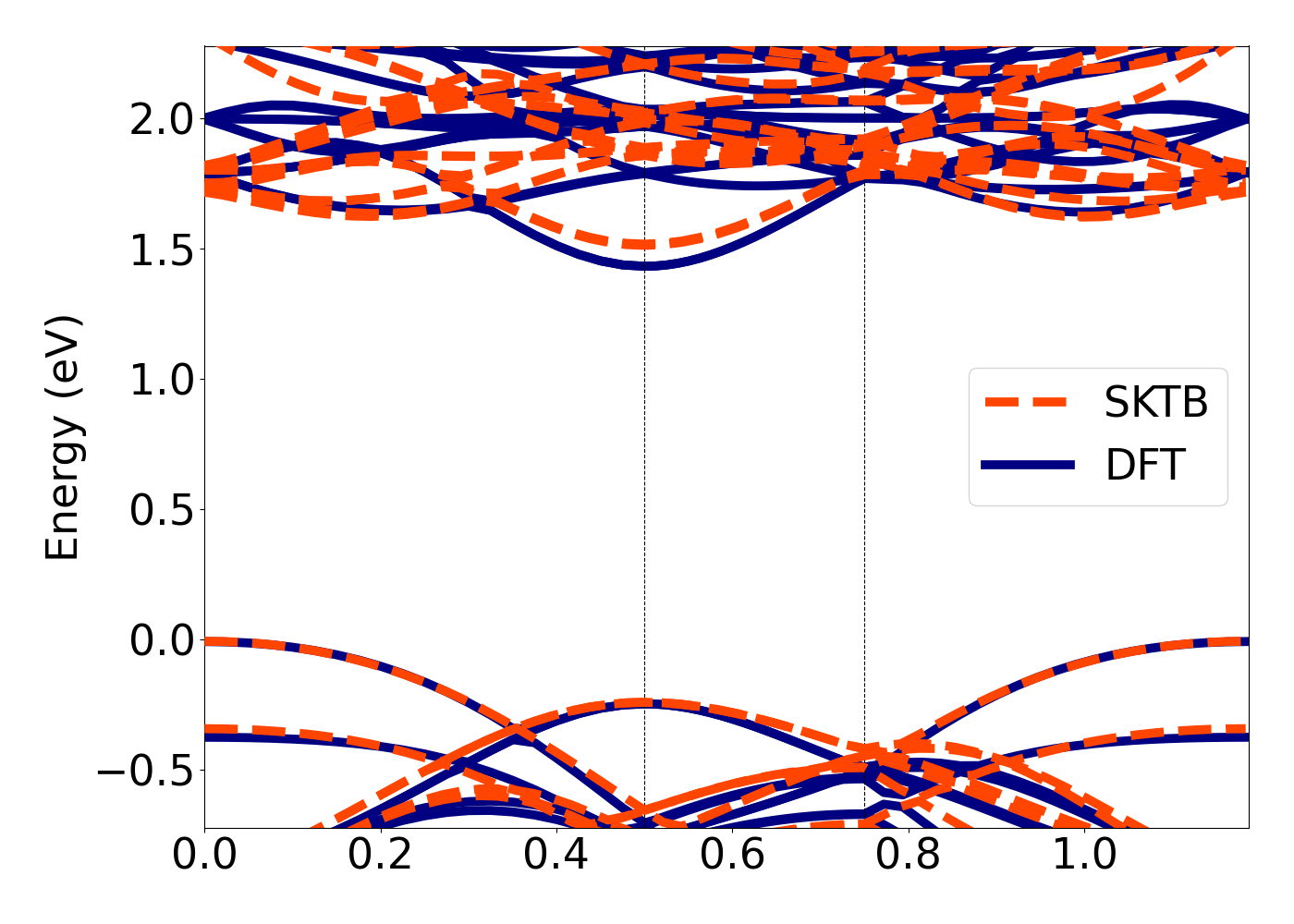}
    \caption{DFT band structure (blue) and SKTB bands (orange) obtained via the parameters in Tab.\ref{tab:intralayer_MoS2} and \ref{tab:interlayer_MoS2} for the smallest commensurate moiré cell ($\theta = 27.787^\circ$)}
    \label{fig:12-bands-moiré}
\end{figure}

As shown in Fig.\ref{fig:12-bands-moiré}, the agreement between SKTB and DFT results is very good, demonstrating that the model can reproduce the electronic structure without any additional fitting parameters, with minimal computational cost. 

As twisted systems clearly break symmetry properties of the AA configuration,  varying the twisting angle allows  new $\sigma_{sc}^{abc}$ components of the current, which we be further tuned. For the particular case of $\theta=21.787^\circ$  the two biggest contributions to the SC conductivity are given by the components: $-\sigma^{xxx}_{sc} = \sigma^{xyy}_{sc} = \sigma^{yxy}_{sc} = \sigma^{yyx}_{sc}$, the same of non twisted system, and the new $-\sigma^{yyy}_{sc} = \sigma^{yxx}_{sc} = \sigma^{xyx}_{sc} = \sigma^{xxy}_{sc}$ components.\\ 

\begin{figure}
    \centering
    \includegraphics[width=1\linewidth]{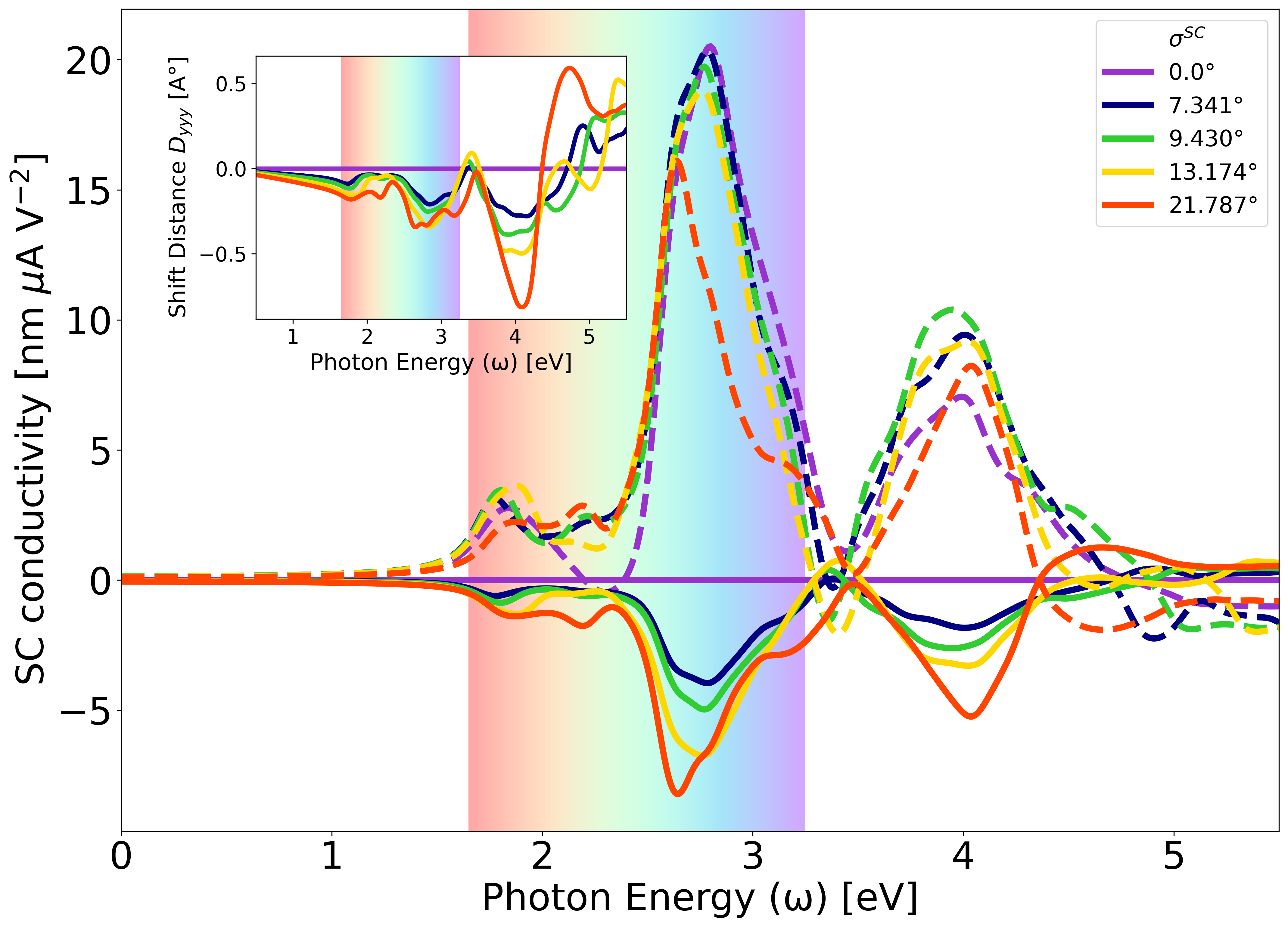}
    \caption{Tensor component $\sigma^{xxx}_{sc}$ (dashed line) and $\sigma^{yyy}_{sc}$ (solid line) of the shift current for four different moiré cells, compared with the untwisted case. In the top left figure the shift-distances, $yyy$ tensor component, for the four twisted configuration and the untwisted, obtained via Eq.\ref{eq:shift-distance}}
    \label{fig:sc-twisted}
\end{figure}

In Fig.\ref{fig:sc-twisted} we report the $\sigma^{xxx}_{sc}$ (dashed lines) and $\sigma^{yyy}_{sc}$ (solid lines) components for four considered twisting angles starting from the non twisted case ($0.0^\circ$) to the largest considered one ($21.787^\circ$). For small angles, the atomic positions of the twisted layer change only slightly with respect to the untwisted system, resulting  in a small variation of the shift current tensor components, $\sigma^{xxx}_{sc}$. On the other hand, even at small angles, $\sigma^{yyy}_{sc}$ are activated due to symmetry breaking, whose contribution strongly increases for larger angles reaching the same order of the major $xxx$ one. This result demonstrates that twisted geometry can indeed be used to activate and modulate the anisotropy of the non-linear shift current, which gain contributions of the same order to $xxx$ and $yyy$ components in MoS$_2$.

The origin of the variation of the conductivity as a function of the angles, can be further analyzed by the calculation of the  shift-distance 
\citep{krishna2023understanding}\citep{nastos2006optical}, which represents the distance, in real space, between the centers of electron charge during the optical absorption, defined as:

\be
D_{ijk} = \frac{\hbar}{\epsilon_0 e}\frac{\sigma_{sc}^{ijk}}{Im(\epsilon^{jk})} 
\label{eq:shift-distance}
–\ee

where 

\be
\epsilon^{jk} = 1 + \frac{i\sigma^{jk}}{\epsilon_0\omega}
\ee

is the dielectric function and $\sigma^{jk}$ the conductivity, $\epsilon_0$ the vacuum permittivity and $e$ the electron charge. In contrast to the shift current, this quantity allows comparison between systems of different volumes because, by construction, it is an intensive property \citep{krishna2023understanding}. 
Observing the $yyy$ component of the shift-distance in Fig.\ref{fig:sc-twisted}, it is evident that in the untwisted case, there is no shift in the distance between the two centers along the $yyy$ component, while in the twisted cases, it exhibits a dependence on the twist angle, increasing up to about 1 \AA~for the larger angle considered.

\begin{figure}
    \centering
    \includegraphics[width=1.0\linewidth]{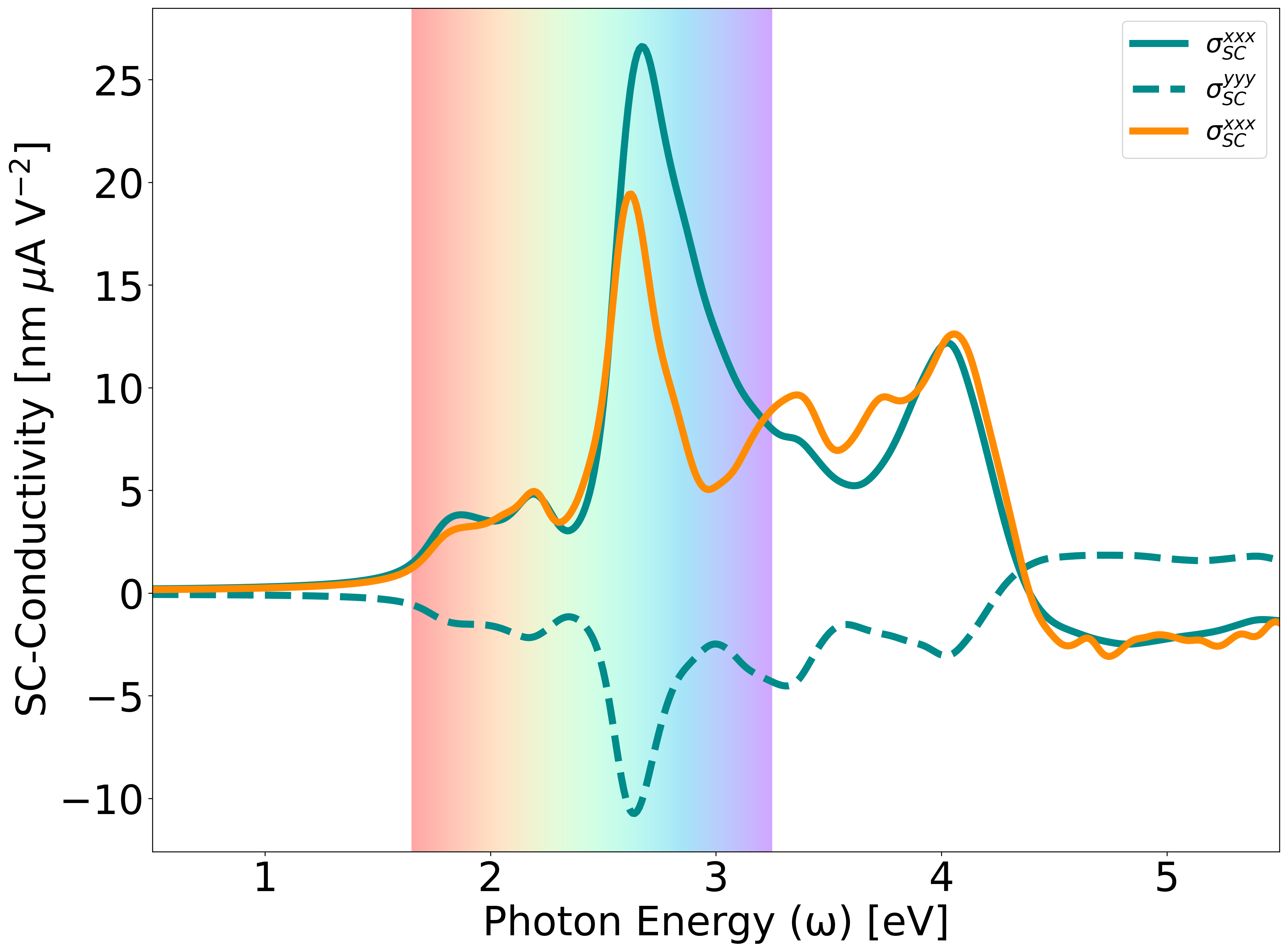}
    \caption{In blue the tensor components $\sigma^{xxx}_{sc}$ and $\sigma^{yyy}_{sc}$ of the shift current for twisted trilayer with an twisting angle of $\theta=21.787^\circ$ for the middle layer. In yellow the tensor components $\sigma^{xxx}_{sc}$ of the shift current for a twisted trilayer where the top and bottom layers are rotated by equal and opposite angles $\pm\theta$, with $\theta = 21.787^\circ$.}
    \label{fig:SC-twisted-trialyers}
\end{figure}

At this point, we can use our TB model to explore even larger systems, considering the twisted geometry in a  MoS$_2$ trilayer. We investigate two different configurations: the first one with only the middle layer twisted by $\theta = 21.787^\circ$, and the second one where the top and bottom layers are rotated by $\pm \theta = 21.787^\circ$ respectively.  
The shift current tensor components $\sigma^{xxx}_{sc}$ and $\sigma^{yyy}_{sc}$ relative to the first geometry considered are shown in Fig.\ref{fig:SC-twisted-trialyers}.
As in the  bilayer case, the shift current exhibits two nonzero orthogonal components, reflecting the reduced symmetry of the system, but we observe an increase both  the $xxx$ and $yyy$ component with respect to the bilayer case, due interlayer interactions which modify the  wavefunctions of the trilayer case.
While for the second configuration we have only one big contribution to the conductivity the -$\sigma^{xxx}_{sc}$  = $\sigma^{xyy}_{sc}$ = $\sigma^{yxy}_{sc}$ = $\sigma^{yyx}_{sc}$, Fig.\ref{fig:SC-twisted-trialyers}, the same non-zero component in the untwisted cases; so the opposite twisting "conserve" the in plane symmetry through the untwisted middle layer.

\quad
\quad

\section{Conclusion}

In this work, we developed and validated a Slater–Koster tight-binding framework for the calculation of the shift-current response in multilayer and twisted MoS$_2$ based systems. By comparing with first-principles calculations for mono-, bi-, and trilayer structures, we demonstrated that the tight-binding model can reproduce the electronic structure near the band edges and captures the main spectral features of the shift-current conductivity. This provides an efficient way for investigating nonlinear optical responses in systems too large for standard density functional theory calculations. 

In multilayer systems, the magnitude of the shift current increases with the number of layers, indicating that interlayer hybridization and the associated redistribution of electronic charge centers can significantly modify the bulk photovoltaic response.
Moreover, we consider the twisting angle as a parameter to break the symmetry. While untwisted structures have only a single set of conductivity tensor elements, the moiré activates additional shift-current conductivity components producing an in-plane anisotropy. The emergence of large ($\sigma^{yyy}$) components and their dependence on twist angle demonstrate that the relative rotation between layers can control both the magnitude and the direction of the photocurrent. The analysis of the shift-distance confirms that the nonlinear response is associated with a modification of the real-space displacement of the centers of charge, providing a direct link between the moiré geometry and the bulk photovoltaic effect. In this sense, twisting modifies the electronic structure through symmetry breaking and directly influences the nonlinear photovoltaic response

More generally, our results establish moiré systems as a promising strategy to control nonlinear photovoltaic phenomena in two-dimensional materials, and call for dedicated experimental verifications. The methodology introduced here can be applied to larger moiré superlattices of twisted van der Waals materials, opening to the possibility of performing large-scale computational calculations for bulk-photovoltaic applications.

Future developments may include the exploration of heterobilayers and correlated moiré systems. Given the strong sensitivity of the shift current to symmetry and wavefunction geometry, twisted transition-metal dichalcogenides and related moiré materials provide a versatile platform for designing tunable nonlinear responses and for new photovoltaic devices.

\section*{Acknowledgements}

C. T. and F. B. acknowledge financial support under the National Recovery and Resilience Plan (NRRP), Mission 4, Component 2, Investment 1.1, funded by the European Union-NextGenerationEU- Project Title “Symmetry-broken HEterostructurEs for Photovoltaic applications-SHEEP”–CUP B53D23028580001-Grant Assignment Decree No. 1409, adopted on 14 September 2022 by the Italian Ministry of Ministry of University and Research (MUR).

\bibliography{bib}

\end{document}